\documentclass[aps,prb,twocolumn,groupedaddress,showpacs,showkeys]{revtex4}
\usepackage{graphicx}
\begin{document}
\title{Shapiro steps in Josephson junctions with alternating critical
current density}
\author{M. Moshe}
\author{R. G. Mints}
\email[]{mints@post.tau.ac.il}

\affiliation{School of Physics and Astronomy, Raymond and Beverly
Sackler Faculty of Exact Sciences, Tel Aviv University, Tel Aviv
69978, Israel}
\date{\today}
\begin{abstract}
We treat theoretically Shapiro steps in tunnel Josephson junctions with
spatially alternating critical current density. Explicit analytical
formulas for the width of the first integer (normal) and half-integer
(anomalous) Shapiro steps are derived for short junctions. We develop
coarse-graining approach, which describes Shapiro steps in the
voltage-current curves of the asymmetric grain boundaries in YBCO thin
films and different superconductor-ferromagnet-superconductor
Josephson-type heterostructures.
\end{abstract}
\maketitle
\section{Introduction}
Series of resonances exist in Josephson tunnel junctions biased at an
alternating voltage
$V(t)=V_0+V_1\cos\omega_1t$.\cite{Shapiro_1,Shapiro_2,Tinkham} In
junctions of conventional superconductors these resonances appear at
the frequencies $\omega_0 =2eV_0/\hbar=n\omega_1$, where $n$ is an
integer. At the resonant values of the dc voltage $V_0
=n\hbar\omega_1/2e$ the supercurrent has a dc component. In the
voltage-current curve Shapiro resonances reveal itself as a ``ladder''
of equidistant values of $V_0$ ({\it integer} Shapiro
steps).\cite{Shapiro_1,Shapiro_2,Tinkham}
\par
The physical origin of Shapiro steps follows from the Josephson
equations $j=j_c\sin\varphi$ and $\dot\varphi =2eV/\hbar$, where $j$
and $j_c$ are the tunneling and critical current densities, and
$\varphi$ is the phase difference across the junction. In order to find
the current across the junction we integrate the voltage $V(t)$ and
find the time-dependent phase $\varphi (t) =\varphi_0 +\omega_0 t +
v_1\sin\omega_1t$. In this relation $\varphi_0 =\varphi (0)$ is the
initial value of the phase and $v_1= 2eV_1/\hbar\omega_1$ is the
dimensionless parameter of the problem. Knowing $\varphi(t)$ we obtain
the tunneling current density in the form $j(t)=j_c\sin(\varphi_0
+\omega_0 t +v_1\sin\omega_1t)$. This formula demonstrates that $j(t)$
is a complex alternating function of time. Fortunately, $j(t)$ can be
transferred into a series allowing for an easy qualitative and
quantitative analysis\cite{Shapiro_2,Tinkham}
\begin{equation}
\label{eq_01}
j=j_c\sum_{n=-\infty}^\infty (-1)^n J_n(v_1) \sin[\varphi_0 + (\omega_0
-n\omega_1)t],
\end{equation}
where $J_n(x)$ is the first kind Bessel function of order $n$.
\par
It is seen from Eq. (\ref{eq_01}) that for any resonant frequency
$\omega_0 =n\omega_1$ the supercurrent density $j$ has a dc component
$\propto\sin\varphi_0$, which reaches its maximum, $j_m$, at $\varphi_0
=\pi/2$. As a result the maximum value of the dc current density is
given by $j_m=j_cJ_n(v_1)\,$.\cite{Shapiro_2,Tinkham}
\par
Anomalous Shapiro steps at the subharmonic resonant frequencies
$\omega_0 =(n/q)\,\omega_1$ ($q>n$, where $q$ and $n$ are integers)
were treated theoretically for short microbridges of conventional
superconductors assuming that the current-phase relation includes
high-order harmonic terms.\cite{Lubbig} Qualitatively the effect of
terms $j_q\sin(q\varphi)$ on the supercurrent density $j$ follows from
Eq. (\ref{eq_01}). Indeed, substituting $\varphi_0$, $\omega_0$ and
$v_1$ by $q\varphi_0$, $q\omega_0$ and $qv_1$ we find that at any
subharmonic frequency $\omega_0=(n/q)\,\omega_1$ the supercurrent
density $j$ has a dc component $\propto\sin q\varphi_0$. In this case
the maximum value of $j$ corresponds to the phase $\varphi_0 =\pi/2q$
and is given by $j_m=j_qJ_n(qv_1)$.
\par
Observation of anomalous half-integer Shapiro steps has been reported
recently for asymmetric grain boundaries in YBCO films\cite{Early} and
superconductor-ferromagnet-superconductor (SFS)
heterostructures.\cite{Sellier_1,Frolov_1,Frolov_2}
\par
%
\begin{figure}
\includegraphics[width=0.95\columnwidth, clip=true]{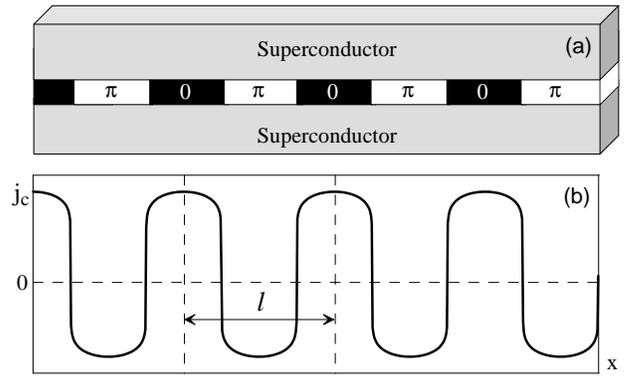}
\caption{Schematic diagrams: (a) Josephson junction made of
periodically interchanging $0\,$- and $\pi\,$-biased fragments; (b)
alternating critical current density distribution.}
\label{fig1}
\end{figure}
%
Qualitative explanations of the origin of the anomalous Shapiro steps
were proposed to understand the results of these experiments mainly by
revealing the existence of second harmonic term, $\sin 2\varphi$, in
the current-phase relation (CPR). It was shown that data measured for
the asymmetric grain boundaries in YBCO are indeed consistent with
existence of the term $\sin2\varphi$ in the CPR.\cite{Schneider,
Golubov, Lindstrom} In the case of SFS junctions, it was assumed that
there is doubling of the Josephson frequency, which leads to a
$\sin2\varphi$ term in the CPR. It was suggested that this frequency
doubling is caused by splitting of energy levels in the ferromagnetic
exchange field.\cite{Sellier_1} An alternative
explanation\cite{Frolov_2} assumes existence of a resonance between the
spontaneous currents and rf modulation.\cite{Vanneste}
\par
Asymmetric grain boundary junctions in YBCO thin films and SFS
heterostructures are arranged in sequences of interchanging $0\,$- and
$\pi\,$-shift tunneling junctions as shown in Fig.~\ref{fig1}.
$\pi\,$-shifts spontaneously appearing across superconducting banks of
tunnel junctions were first considered for SFS
heterostructures.\cite{Bulaevskii, Buzdin} These $\pi\,$-shifts lead to
negatively biased critical current density, $j_c$, and anomalous
Josephson properties of sequences of $0\,$-$\,\pi$ biased tunnel
junctions.\cite{Zenchuk, Ryazanov, Goldobin, Sellier_2, Radovic,
Hilgenkamp_1} The observed anomalies are especially significant if the
lengths of $0\,$- and $\pi\,$-fragments are much less than the
Josephson penetration depth $\lambda_J$ defined by the spatial average
of $|j_c|$.\cite{Mannhart, Hilgenkamp_2, Mints_1, Mints_2}
\par
Thus, several qualitative explanations have been put forward to
understand the physics of fractional Shapiro steps. However, to the
best of our knowledge there is at present no theoretical model
describing the anomalous Shapiro steps and their basic features in
Josephson junctions with alternating critical current density.
\par
In this paper we treat Shapiro steps in Josephson junctions arranged in
periodic or almost periodic sequences of interchanging $0\,$- and
$\pi\,$-biased fragments with $l\ll\lambda_J$. We develop a
coarse-graining approach which is applicable to treat time-dependent
phenomena.
\par
In our theoretical model we assume that the tunneling current density
is given by the {\it standard} CPR dependence
$j=j_c(x)\sin\varphi(x,t)$, but with a critical current density,
$j_c(x)$, spatially alternating along the Josephson junction. As an
illustration we treat the case of a short junction that allows to
obtain explicit analytical results. We demonstrate that in Josephson
junctions with spatially alternating critical current density the
half-integer (anomalous) Shapiro steps exist in addition to the integer
(normal) Shapiro steps. We calculate the dependence of the width of
these steps on the flux inside the junction and the voltage across the
junction.
\par
The organization of this paper is as follows. In Sec.~II we review
briefly the coarse-graining approach to Josephson junctions with
spatially alternating critical density. In Sec.~III this general
approach is applied to calculate explicit formulae for the width of the
integer (normal) and half-integer (anomalous) Shapiro steps in short
junctions. Sec.~IV discusses these results.
\par
\section{Josephson junction with spatially alternating critical current
density}
Consider a tunnel junction of length $L$ ($0\leq x\leq L$) with a
critical current density $j_c(x)$ alternating with a length-scale
$l\gg\lambda$, where $\lambda$ is the London penetration depth (see
Fig. \ref{fig1} for the geometry of the problem). It is convenient for
the following calculations to write
\begin{equation}\label{eq_02}
j_c(x)=\langle j_c\rangle [1+g(x)]\,,
\end{equation}
where $\langle j_c\rangle$ is the average value of $j_c(x)$ defined as
\begin{equation}\label{eq_03}
\langle f\rangle ={1\over L}\,\int_0^L \!\!\! dx\,f(x)\,.
\end{equation}
The function $g(x)$ describes spatial variations of $j_c(x)$ and has a
zero average value, $\langle g(x)\rangle =0$. In what follows we are
interested in the case of small average values of the critical current
density, {\it i.e.}, we assume that $\langle j_c(x)\rangle\ll\langle
|j_c(x)|\rangle$. This condition means that the typical value of the
dimensionless function $g(x)$ is big compared to unity or in other
words $\langle|g(x)|\rangle\gg 1$.
\par
In the above notations equation for the phase difference $\varphi(x,t)$
across the junction reads
\begin{equation}\label{eq_04}
\tau^2\ddot{\varphi}-\Lambda^2\varphi''+
\left[1+g(x)\right]\sin\varphi=0\,,
\end{equation}
where $1/\tau$ is the Josephson frequency, and $\Lambda$ is the
effective Josephson penetration depth,
\begin{equation}\label{eq_05}
\Lambda^2={c\phi_0\over16\pi^2\lambda\langle j_c\rangle}\,.
\end{equation}
\par
We treat now a sample subjected to magnetic field $H_a$ and alternating
voltage $V(t)=V_0+V_1\cos\omega_1t$ applied across the junctions. It is
assumed that $2eV_1$ is small compared to $\hbar\omega_1$, {\it i.e.},
$v_1=2eV_1/\hbar\omega_1\ll 1$. In this case coarse-graining can be
applied to solve Eq. (\ref{eq_04}).\cite{Mints_1}
\par
Two types of terms appear in Eq. (\ref{eq_04}): terms alternating over
the length $l$ and smooth terms varying over the length $\Lambda\gg l$.
The fast alternating terms cancel each other, independently of the
smooth terms, which also cancel each other. Therefore, we
write\cite{Mints_1}
\begin{equation}
\label{eq_06}
\varphi (x)=\psi (x)+\xi (x)\,,
\end{equation}
where $\psi (x)$ is a smooth function with the length-scale of order
$\Lambda$ and $\xi (x)$ alternates with the length-scale of order $l$.
Under the above assumptions the average value of $\xi (x)$ is zero and
the typical amplitude of variations of $\xi (x)$ is small, {\it i.e.},
$\langle\xi(x)\rangle =0$ and $\langle|\xi(x)|\rangle\ll 1$.
\cite{Mints_1,Mints_2,Landau_1,Arnold_1}
\par
Substituting Eq.~(\ref{eq_06}) into Eq.~(\ref{eq_04}) and keeping terms
up to first order in $\xi (x)$ we find\cite{Mints_1}
\begin{eqnarray}
\label{eq_07}
\Lambda^2\psi'' - {j_\psi (x)\over\langle j_c\rangle} &=& 0\,,\\
\label{eq_08}
\Lambda^2\xi'' - {j_\xi (x)\over\langle j_c\rangle} &=& 0\,,
\end{eqnarray}
where the smooth $j_\psi (x)$ and alternating $j_\xi (x)$ components of
the tunneling current density $j= j_\psi +j_\xi$ are
\begin{eqnarray}
\label{eq_09}
j_\psi &=& \langle j_c\rangle \left(\sin\psi -
\gamma\sin\psi\cos\psi\right)\,,\\
\label{eq_10}
j_\xi &=& \langle j_c\rangle\,g(x)\sin\psi\,.
\end{eqnarray}
The dimensionless constant $\gamma$ is equal to
\begin{equation}
\label{eq_11}
\gamma = \langle g(x)\xi_g (x)\rangle =
\Lambda^2\langle \xi_g^{\prime 2}(x)\rangle >0\,,
\end{equation}
and the rapidly alternating phase $\xi_g(x)$ is defined by
\begin{equation}
\label{eq_12}
\xi (x) = -\xi_g (x)\sin\psi\,.
\end{equation}
It follows from Eqs. (\ref{eq_08}), (\ref{eq_10}) and (\ref{eq_12})
that
\begin{equation}
\label{eq_13}
\Lambda^2\xi''_g + g(x) = 0\,,
\end{equation}
{\it i.e.}, the rapidly alternating phase shift $\xi_g(x)$ is a
characteristics of a sample.
\par
To summarize the coarse-graining approach it is worth noting that the
typical values of $\xi_g(x)$ are small, but at the same time the
typical values of $g(x)$ are big, {\it i.e.},
$\langle|\xi_g(x)|\rangle\ll 1$ and $\langle|g(x)|\rangle\gg 1$. As a
result, the dimensionless parameter $\gamma$, which is proportional to
the average of the product of the two rapidly alternating functions
$\xi_g(x)$ and $g(x)$ might be of the order of unity.
\cite{Mints_1,Mints_2} The value of $\xi'_g$ can be estimated as
$\xi'_g\sim \xi_g/l$. It follows then from Eq. (\ref{eq_11}) that
$\langle\,\xi^2_g\rangle\sim\gamma l^2/\Lambda^2\ll 1$. Therefore for
$\xi'_g$ we have the estimate $\Lambda\langle|\xi'_g|\rangle
\sim\sqrt{\gamma}\sim 1$.
\par
\section{Shapiro steps in short junctions}
In a short junction ($L\ll\Lambda$) $\psi (x,t)$ is almost linear in
$x$, the time dependence of $\psi(x,t)$ is given by $V(t)$, and
\begin{eqnarray}
\label{eq_14}
&&\psi'(x,t) ={4\pi\lambda\over\phi_0}\,H_i(t)\,,\\
\label{eq_15}
&&\dot{\psi}(L/2,t) ={2e\over\hbar}\,V(t)\,,
\end{eqnarray}
where $H_i(t)$ is the time-dependent field in the junction.
\par
At the sample edges the derivative $\varphi' =\psi' +\xi'$ is
proportional to the field $H|_{0,L}$ including the self-field generated
by the total current $I$.\cite{Barone_1} In the case of a short
junction we obtain
\begin{equation}
\label{eq_16}
\left({H_a\over H_s}\mp{1\over2}{I\over I_s}\right)_{\rm 0,L}
=\Lambda\varphi'\big|_{\rm 0,L} =\Lambda\psi'\big|_{\rm 0,L}
+\Lambda\xi'\big|_{\rm 0,L}\,,
\end{equation}
where $H_s=\phi_0/4\pi\lambda\Lambda$, $I_s=\langle j_c\rangle\Lambda$.
It follows from Eq. (\ref{eq_12}) that the derivative
$\xi'=-\xi'_g\sin\psi$. In order to find $\xi'_g$ we integrate Eq.
(\ref{eq_13}) from $0$ to $L$ and arrive to the relation $\xi'_g(0)
=\xi'_g(L)$, which allows to rewrite the boundary condition
(\ref{eq_16}) in the final form
\begin{equation}
\label{eq_17}
\left({H_a\over H_s}\mp{1\over 2}{I\over I_s}\right)_{\rm 0,L}
=\Lambda\psi'\big|_{\rm 0,L} -\alpha\,\sin\psi\big|_{\rm 0,L}\,,
\end{equation}
where the terms $\alpha\sin\psi\big|_{\rm 0,L}$ are caused by the
high-density edge currents and $\alpha$ is a constant. It is worth
noting here that $\alpha$ is a characteristics of a sample. Next, using
the above estimate for $\xi'_g$ we obtain
\begin{equation}
\label{eq_18}
\alpha =\Lambda\xi_g'\big|_{\rm 0,L}\sim\sqrt{\gamma}\sim 1.
\end{equation}
\par
To summarize the above analysis, we find that in the coarse-graining
approach the Josephson junctions with spatially alternating critical
current density and $l\ll\lambda_J$ are characterized by two
dimensionless parameters $\gamma$ (for the inner part of the junction)
and $\alpha$ (for the edges of the junction), where
$\lambda_J=\sqrt{c\phi_0/16\pi^2\lambda\langle |j_c|\rangle}\ll\Lambda$
is the local Josephson length.
\par
Next, we find the derivatives $\psi'\big|_{\rm 0,L}$ using the first
integral of Eq. (\ref{eq_07}), that can be written as
\begin{equation}
\label{eq_19}
{\Lambda^2\over 2}\psi'^2 + \cos\psi -{\gamma\over 4}\cos 2\psi = {\rm
Const}.
\end{equation}
Finally, we combine Eqs. (\ref{eq_17}) and (\ref{eq_19}) relating the
total current $I$, applied field $H_a$, and phase $\psi$ at the edges.
In the case of a short junction we find
\begin{eqnarray}
\label{eq_20}
&&I=2I_s\sin\left(\pi\,{\phi_i\over\phi_0}\right)\bigg\{\alpha\,\cos\psi_m -
\nonumber\\
&&-{L\over 2\pi\Lambda}{\phi_0\over\phi_a}
\left[1-\gamma\cos\psi_m\cos\left(\pi\,{\phi_a\over\phi_0}\right)\right]
\sin\psi_m\bigg\},\\
\label{eq_21}
&&\phi_i=\phi_a+{\alpha\over 2\pi}\,{L\over\Lambda}\,
\phi_0\cos\left(\pi{\phi_a\over\phi_0}\right)\sin\psi_m\,,\\
\label{eq_22}
&&\psi_m=\psi(L/2,t)=\psi_0 +\omega_0 t + v_1\sin\omega_1 t,
\end{eqnarray}
where the ``applied'', $\phi_a$, and ``internal'', $\phi_i$, fluxes are
defined as $\phi_{a,i}=2L\lambda H_{a,i}$, $\omega_0 =2eV_0/\hbar$, and
$\psi_0$ is a constant, which is used to maximize the total current
$I$.
\par
Calculation of the width of Shapiro steps similar to the one given by
M. Tinkham\cite{Tinkham} leads to the following results: two series of
steps appear at frequencies
\begin{equation}
\label{eq_23}
\omega_0=(n+1/2)\,\omega_1\quad {\rm and} \quad
\omega_0=n\,\omega_1,
\end{equation}
where $n$ is an integer.
\par
In the case of low oscillating voltage ($v_1\ll 1$) we obtain the
widths of the first {\it half-integer}, $I_{1\over 2}(\phi_a)$, and the
first {\it integer}, $I_1(\phi_a)$, Shapiro steps in the form
\begin{equation}
\label{eq_24}
I_{1\over 2}=
v_1I_c\left|\alpha^2\cos^2\left(\pi{\phi_a\over\phi_0}\right)
+{\gamma\phi_0\over2\pi\phi_a}
\sin\left(2\pi{\phi_a\over\phi_0}\right)\right|\,,
\end{equation}
\begin{equation}
\label{eq_25}
I_1=v_1I_c
\sqrt{\left({\alpha\Lambda\over L}\right)^2 +
\left({\phi_0\over\pi\phi_a}\right)^2 }
\left|\,\sin\left(\pi{\phi_a\over\phi_0}\right)\right|\,,
\end{equation}
where $I_c= \langle j_c\rangle L$. Explicit formulas (\ref{eq_24}) and
(\ref{eq_25}) reveal quite a few remarkable features of $I_{1\over
2}(\phi_a)$ and $I_1(\phi_a)$. In what follows we discuss them in
details.
\par
\begin{figure}
\includegraphics[width=0.95\columnwidth]{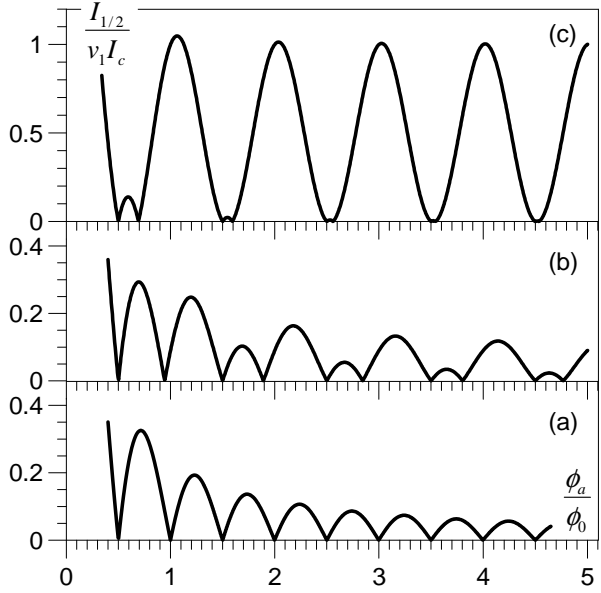}
\caption{Dependence of the width of the first half-integer Shapiro step,
$I_{1\over 2}$, on applied flux, $\phi_a$, given by Eq. (\ref{eq_24})
for $\gamma = 1.5$ and (a) $\alpha =0.03$, (b) $\alpha =0.3$, and (c)
$\alpha =1$.}
\label{fig2}
\end{figure}
\par
\section{Results and discussion}
The function $I_{1\over 2}(\phi_a)$ is equal to zero for the {\it two}
series: $\phi_a =\left(n+{1/2}\right)\phi_0$ and $\phi_a = (n -
\psi_\theta /\pi)\phi_0$, where $n$ is an integer and $\psi_\theta$ is
defined by $\tan\psi_\theta =\pi\alpha^2\phi_a/\gamma\phi_0$. The angle
$\psi_\theta$ depends on $\phi_a$, {\it i.e.}, in general, the roots of
equation $I_{1\over 2} (\phi_a)=0$ are not equidistant and the {\it
actually observed} dependence $I_{1\over 2}(\phi_a)$ is defined by the
dimensionless ratio $\psi_a =\gamma/2\pi\alpha^2$.
\par
Next, it follows from Eq. (\ref{eq_24}) that for $\phi_a=0$ the width
of the first anomalous half-integer step $I_{1\over 2}(0)$ equals to
$v_1I_c(\gamma +\alpha^2)$. In the {\it low-flux} region
($\phi_a/\phi_0\ll\psi_a$) the main contribution to $I_{1\over
2}(\phi_a)$ comes from the second term in Eq. (\ref{eq_24}). This term
originates from the alternating currents flowing across {\it all}
junctions. As a result, in the low-flux region we have $I_{1\over
2}\propto (\phi_0/2\pi\phi_a) |\sin(2\pi\phi_a/\phi_0)|$, {\it i.e.},
$I_{1\over2}(\phi_a)$ is described by the Fraunhofer pattern but with a
double frequency. In Fig. \ref{fig2}\,(a) we show $I_{1\over
2}(\phi_a)$ in the low-flux region ($\phi_a/\phi_0\ll 250$) for the
data $\alpha =0.03$ and $\gamma =1.5$, which lead to $\psi_a\approx
250$.
\par
In the {\it high-flux} region ($\phi_a/\phi_0\gg\psi_a$) the main
contribution to $I_{1\over 2}$ comes from the first term in Eq.
(\ref{eq_24}). This term originates from the high density currents
flowing across the {\it edge} junctions. As a result, in the high-flux
region we have $I_{1\over 2}\propto
\cos^2(\pi\phi_a/\phi_0)$, {\it i.e.}, $I_{1\over 2}(\phi_a)$ is a
periodic function. In Fig. \ref{fig2}\,(c) we show $I_{1\over 2}
(\phi_a)$ in the high-flux region ($\phi_a/\phi_0\gg 0.25$) for the
data $\alpha =1$ and $\gamma =1.5$, which lead to $\psi_a\approx 0.25$.
\par
In Fig. \ref{fig2}\,(b) we show the function $I_{1\over 2}(\phi_a)$ for
the intermediate values of the applied flux
($\phi_a/\phi_0\sim\psi_a$). Using the data $\gamma =1.5$ and $\alpha
=0.3$, we find that $\psi_a\approx 2.5$. In this case $\psi_\theta$ is
strongly flux dependent. As a result the function $I_{1\over
2}(\phi_a/\phi_0)$ is manifestly aperiodic.
\par
We discuss now the width of the first integer Shapiro step, $I_1$. It
follows from Eq. (\ref{eq_25}) that $I_1(\phi_a)$ is equal to zero for
the series: $\phi_a =n\phi_0$, where $n$ is an integer.
\par
\begin{figure}
\includegraphics[width=0.95\columnwidth]{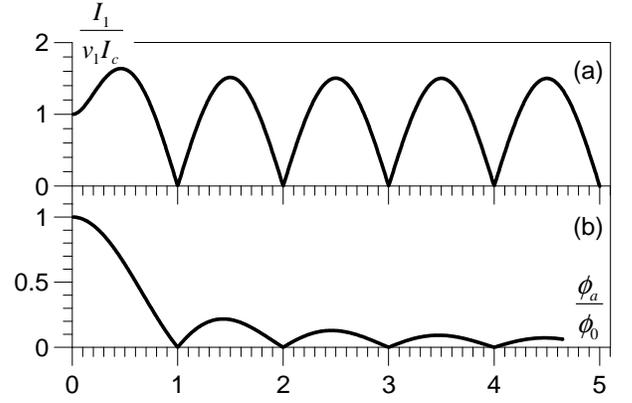}
\caption{Dependence of the width of the first integer Shapiro step,
$I_1$, on applied flux, $\phi_a$, given by Eq. (\ref{eq_25}) for
$L/\Lambda = 0.2$, and (a) $\alpha =0.3$, and (b) $\alpha = 0.003$.}
\label{fig3}
\end{figure}

In the {\it high-flux} region ($\phi_a/\phi_0 \gg L/\pi\alpha\Lambda$)
the main contribution to $I_1$ comes from the first term in Eq.
(\ref{eq_25}). This term originates from the high density currents
flowing across the {\it edge} junctions. As a result, in the high-flux
region we obtain $I_1(\phi_a)\propto |\sin(\pi\phi_a/\phi_0)|$, {\it
i.e.}, $I_1(\phi_a)$ is a periodic function.  In Fig. \ref{fig3}\,(a)
we show $I_1(\phi_a)$ in the high-flux region ($\phi_a/\phi_0\gg 0.2$)
for the data $\alpha =0.3$ and $L/\Lambda =0.2$, which lead to
$L/\pi\alpha\Lambda\approx 0.2$.
\par
Next, it follows from Eq. (\ref{eq_25}) that the width of the first
integer step $I_1(0)$ equals to $v_1I_c$. In the {\it low-flux} region
($\phi_a/\phi_0\ll L /\pi\alpha\Lambda$) the main contribution to $I_1$
comes from the second term in Eq. (\ref{eq_25}). This term originates
from the alternating currents flowing across {\it all} junctions. As a
result, in the low-flux region we obtain $I_{1\over 2}\propto
(\phi_0/\pi\phi_a)\sin(\pi\phi_a/\phi_0)$, {\it i.e.}, $I_1(\phi_a)$ is
described by the Fraunhofer pattern. In Fig. \ref{fig3}\,(b) we show
$I_1(\phi_a)$ in the low-flux region ($\phi_a/\phi_0\ll 20$) for the
data $\alpha =0.003$ and $L/\Lambda =0.2$, which lead to
$L/\pi\alpha\Lambda\approx 20$.
\par
\par
Let us now illustrate the above calculations by using a model
dependence for alternating critical current density
\begin{equation}
\label{eq_26}
j_c(x)=j_0+j_1\sin\left({2\pi\over l}\,x +\theta\right),
\end{equation}
where $j_0=\langle j_c(x)\rangle$ and $j_1$ are constants, $\theta$ is
an angle from the interval $0\leq\theta\leq\pi$, and $L/l=N$ is an
integer ($N\gg 1$). It follows from Eqs. (\ref{eq_02}) and
(\ref{eq_13}) that
\begin{eqnarray}
\label{eq_27}
g(x)&=&{j_1\over j_0}\sin\left({2\pi\over l}\,x +\theta\right),\\
\label{eq_28}
\xi_g(x)&=&{4\lambda l^2j_0\over c\phi_0}\, g(x)\,.
\end{eqnarray}
Next, knowing $g(x)$ and $\xi_g(x)$ we use Eqs. (\ref{eq_11}) and
(\ref{eq_20}) and find the parameters
\begin{eqnarray}
\label{eq_29}
\gamma &=& {2\lambda l^2j_1^2\over c\phi_0j_0},\\
\label{eq_30}
\alpha &=&\sqrt{2\gamma}\cos\theta\,.
\end{eqnarray}
\par
It follows from Eqs. (\ref{eq_24}) and (\ref{eq_25}) that the widths of
the first half-integer $I_{1\over 2}$ and integer $I_1$ Shapiro steps
are
\begin{equation}
\label{eq_31}
I_{1\over 2}={v_1j_1L\over\cos\psi_\theta }\,{l_1^2\over\Lambda_1^2}
\left|\cos\left({\pi\phi_a\over\phi_0}\right)
\cos\left({\pi\phi_a\over\phi_0}-\psi_\theta\right)\right|,
\end{equation}
\begin{equation}
\label{eq_32}
I_1=v_1\sqrt{\left(j_0L{\phi_0\over\pi\phi_a}\right)^2
+\left({j_1l_1\over4\pi}\right)^2}
\left|\sin\left({\pi\phi_a\over\phi_0}\right)\right|,
\end{equation}
where we define $\tan\psi_\theta=(\phi_0/2\pi\phi_a)/\cos^2\theta$,
$l_1=l\cos\theta$, and $\Lambda_1^2 ={c\phi_0/4\lambda j_1}$.
\par
An interesting feature follows from Eq.(\ref{eq_32}) for the limiting
case $j_0=0$. Using Eq.(\ref{eq_32}) we find that if the average value
of the critical current density $j_0=0$, then the width of the first
integer Shapiro step equals to
\begin{equation}
\label{eq_33}
\tilde{I}_{1}(\phi_a)={2eV_1\over\hbar\omega_1}\, {j_1l\over 2\pi}\,
\cos\theta\,\left|\sin\left({\pi\phi_a\over\phi_0}\right)\right|.
\end{equation}
This current is the contribution coming from the edge fragments and
therefore the value of $\tilde{I_1}$ does not depend on the total
length of the junction $L$.
\par
\section{Summary}
The anomalous Shapiro steps in our model exist due to: (a) successful
interference between the spatial alternations of the critical current
density $j_c(x)$ and phase factor $\sin\varphi (x,t)$, leading to
generation of the second harmonic in the Josephson current density; (b)
high current density at the edges of the junction resulting in
anomalous dependencies $I_{\rm 1\over 2}(H_a)$ in terms of periodicity
at low fields and asymptotic behavior at high fields.
\par
To summarize, we demonstrate the existence of anomalous half-integer
Shapiro steps for short Josephson junctions with spatially alternating
critical current density $j_c(x)$. We derive explicit formulas given by
Eqs. (\ref{eq_24}) and (\ref{eq_25}) for the width of the first integer
and half-integer Shapiro steps. The general approach is applied to the
case of a simple model for the critical current density dependence on
the coordinate along the junction (see Eq. (\ref{eq_26})). The results
obtained in the framework of this model might be useful for analysis of
the experimental data for the asymmetric grain boundaries in YBCO thin
films and different superconductor-ferromagnet-superconductor
Josephson-type heterostructures.
\par
\begin{acknowledgments}
One of the authors (RGM) is grateful to J. R. Clem, V. G. Kogan, J.
Mannhart, and C. W. Schneider for support and numerous stimulating
discussions.
\end{acknowledgments}
\par
\end{document}